\documentclass[prb,twocolumn,preprintnumbers,superscriptaddress]{revtex4-2}

\usepackage{graphicx} 

\begin{document}

\title{Effect of two-dimensional nonlocal screening on mobility of electrons \\ in transition-metal dichalcogenide monolayers}

\author{Aram Manaselyan}
\email{amanasel@ysu.am}
\affiliation{Department of Condensed Matter Physics, Yerevan State University, Alex Manoogian 1, 0025 Yerevan, Armenia}

\author{Vram Mughnetsyan}
\email{vram@ysu.am}
\affiliation{Department of Condensed Matter Physics, Yerevan State University, Alex Manoogian 1, 0025 Yerevan, Armenia}

\author{Anna Asatryan}
\email{annaa@ysu.am}
\affiliation{Department of Condensed Matter Physics, Yerevan State University, Alex Manoogian 1, 0025 Yerevan, Armenia}

\author{Albert Kirakosyan}
\email{kirakosyan@ysu.am}
\affiliation{Department of Condensed Matter Physics, Yerevan State University, Alex Manoogian 1, 0025 Yerevan, Armenia}

\begin{abstract}
A new mechanism for charge carrier scattering in transition-metal dichalcogenide monolayers is proposed on the basis of the theory of two-dimensional nonlocal screening developed for the dielectric function of thin-layer insulating materials (P. Cudazzo et al. PRB 84, 085406 (2011)). The expressions for the transport relaxation time and for the electron mobility are obtained for electrons scattering on Coulomb impurity centers in monolayers of transition-metal dichalcogenide on various substrates. It is found that taking nonlocal screening into account increases the mobility of electrons by several times. Although the value of the mobility decreases with increasing temperature, the relative enhancement due to nonlocal screening grows 6--9 times at room temperature, in the case of SiO$_2$ substrate.

\noindent\textbf{Keywords:} transition metal dichalcogenides, transport relaxation time, mobility, Coulomb scattering, nonlocal screening
\end{abstract}

\maketitle

\section{Introduction}
Studies of low-dimensional systems represent the frontiers of modern condensed matter physics both in terms of fundamental physics \cite{Marino} and its practical applications \cite{Wang,Qiuyang}, as well as in terms of finding new nanomaterials and modern methods for their synthesis \cite{Pacile, Bonaccorso, Novoselov, Sofo, Elias, Chao, Herath, Romera, Vigneshwaran}. In particular, two-dimensional (2D) materials have attracted great attention after the discovery of single layer graphene that has exceptional physical properties \cite{Castro-Neto, Geim}.
Transition-metal dechalcogenides (TMDs) are 2D semiconductors exhibiting a unique combination of atomic-scale thickness and strong spin-orbit coupling, which makes them promising candidates for spintronics \cite{Zhong, Luo, Avsar}, primarily direct band gaps which are of great interest for applications in photonics and optoelectronics \cite{Mak, Yin}. 
Semiconducting 2D TMDs have unique features that make them attractive as channel material in field effect transistors (FETs), such as the lack of dangling bounds, structural stability, and mobility comparable to Si \cite{Fivaz, Podzorov}.

The idea of using TMDs for transistor applications was proposed in \cite{Novoselov, Podzorov}. In \cite{Radisavljevic} the authors reported higher performance monolayer $\text{MoS}_2$ target transistors. Since then, enormous progress has been made to understand, particularly charge transport in TMDs. It should be noted the reference \cite{Li2016} which is a review of progress on charge transport properties and carrier mobility engineering of 2D TMDs, with particular focus on the markedly high dependence on carrier mobility on thickness. 
The mobility of charge carriers (CC) in the FET channel is formed under the influence of various scattering mechanisms, as well as the structure and material of the channel and the properties of the environments surrounding it.

In \cite{Li2013, Li2015} the combined experimental and theoretical studies of the origins of the dependence of carrier mobility on thickness of the specimen are performed and revealed that the expended injection barrier at contacts with the decreasing thickness and interfacial Coulomb impurities are the main factors responsible for the observed thickness dependence. \cite{Ju2022} is devoted to experimental and theoretical studies on the fundamental Coulomb screening and scattering effects in 2D TMDs layered systems. \cite{Zhang2023} is a comprehensive theoretical survey on the performance limits for some monolayer $\text{MoS}_2$ transistors by including primary extrinsic charge scattering mechanisms present in practical devices. The charge mobility and current density are analyzed for transistors at post-silicon technological nodes beyond 1nm. In \cite{Liang2023} the feasibility of enhancing carrier mobility in 2D semiconductors is shown through engineering the vertical distribution of carriers confined inside ultrathin channels via symmetrizing gate configuration or increasing channel thickness. The scattering mechanism responsible for limiting the mobility of single layer semiconductors is evaluated in \cite{Ma2014}. In \cite{Yu2017} taking $\text{MoS}_2$ as an example, the key factors that reduce mobility in TMDs transistors are reviewed. A theoretical model that quantitatively captures the scaling of mobility with temperature, carrier density and thickness is introduced. In \cite{Mu2024} authors developed a comprehensive theoretical model to decouple experimental mobilities of monolayer $\text{MoS}_2$ transistors, which in turn provides insight into the electron transport mechanism in different systems. The model accurately fits various electron transport mechanisms for monolayer $\text{MoS}_2$ transistors, particularly those dominated by lattice phonon scattering or dominated localized charge trap effects.

The effect of semiconducting layer thickness and dielectric constants of both the layer and surrounding media on the Coulomb interaction of charge carriers in quasi-2D systems was first considered in \cite{Rytova, Keldysh}. However for the atomic-scale semiconducting layers consisting of a few monolayers down to a monolayer the results obtained in \cite{Rytova, Keldysh} contradict with the meaning of the macroscopic dielectric constant of the layer.
In \cite{Cudazzo1} for atomic thin layer of insulating material an exact analytic form of the 2D screened potential is derived, and it is shown that in 2D systems the macroscopic screening is nonlocal, so that in the Fourier space it is described by a wave-number-dependent macroscopic dielectric function. In the frame of the developed theory it was particularly shown that the impurity hole doping in graphane leads to strongly localized states. In \cite{Cudazzo2} the estimation of the bound energy of exciton is made by the variational solution of Schrodinger equation with effective Coulomb potential and the coincidence of theoretical and experimental values is explained by the nonlocal character of 2D dielectric screening.
On the bases of 2D-screening theory developed in \cite{Cudazzo1}, in \cite{Berkelbach} the microscopic theory of neutral and charged excitons in a TMD monolayer is presented and in \cite{Durnev} the theoretical and experimental results of excitonic effects in monolayers of TMDs are reviewed. In the frame of variational method in \cite{Mughnetsyan} the expressions of the energy and the effective radius of the impurity ground state depending on the effective screening parameter of the problem are obtained both without and with magnetic field.
In \cite{Zhun} the charged impurity scattering and static screening in top-gated single layer graphene is studied.

The aim of this work is to determine the influence of nonlocal screening \cite{Cudazzo1} on the kinetic characteristics of TMD monolayers, in particular, on the transport relaxation time and the mobility of electrons in monolayers located between dielectric media with different dielectric constants. To the best of our knowledge, no theoretical study on the effect of nonlocal screening on the kinetic characteristics of TMD monolayers has been reported elsewere. 

The paper is organized as follows: in section II the theoretical model is described, section III is devoted to the results analyses and discussions, and in section IV the concluded remarks are included.

\section{Theoretical framework}
\begin{table*}
\caption{\label{tab:tmd_parameters}
Lattice constants \cite{Ashwin}, reduced electron masses \cite{Rybkovskiy}, characteristic lengths \cite{Berkelbach}, effective Bohr radii and the parameter of the problem in monolayer TMDs.
}
\begin{ruledtabular}
\begin{tabular}{lcccccccccc}
TMD & $a$ (\AA) & $m/m_0$ & $r_0$ (\AA) & \multicolumn{3}{c}{$a_B$ (\AA)} & \multicolumn{3}{c}{$\delta$} \\
\cline{5-7} \cline{8-10}
& & & & SiO$_2$ & Al$_2$O$_3$ & HfO$_2$ & SiO$_2$ & Al$_2$O$_3$ & HfO$_2$ \\
\hline
MoS$_2$   & 3.18 & 0.37 & 41.45 & 3.50 & 9.68 & 16.44 & 4.83 & 0.63 & 0.22 \\
MoSe$_2$  & 3.32 & 0.52 & 51.68 & 2.50 & 6.89 & 11.70 & 8.46 & 1.11 & 0.38 \\
WS$_2$    & 3.19 & 0.40 & 37.38 & 3.24 & 8.95 & 15.21 & 4.77 & 0.62 & 0.22 \\
WSe$_2$   & 3.39 & 0.46 & 45.09 & 2.82 & 7.79 & 13.23 & 6.53 & 0.86 & 0.30 \\
\end{tabular}
\end{ruledtabular}
\end{table*}
In 2D TMDs layers transport and scattering of the carriers are confined to the plane of the specimen. The mobility of carriers in low-dimensional systems is affected by the main scattering mechanisms: Coulomb scattering \cite{Li2013, Li2015, Ju2022, Liang2023, Yu2017, Mu2024}, phonon scattering \cite{Kaasbjerg, Kaasbjerg2013, Yu2017, Liang2023, Konar2010, Zeng2013, Li2016}  including the remotely polar-optical phonon modes being excited in surrounding dielectrics \cite{Konar2010, Fratini2008, Fischetti2001}, scattering by structural defects \cite{Zhang2023, Mu2024} and scattering by surface roughnesses \cite{Ando}.
However, it should be noted that for a monolayer of TMD the latter scattering mechanism can apparently be neglected.
Indeed, the ratio of the electron wavelength $\lambda$ to the lattice constant $a$ (see Table I) at the density of 2D electrons
$n \leq n_{max}=10^{14} cm^{-2}$, $\lambda / a \geq 10$, which means the specular nature of reflection from the surfaces of the monolayer. Moreover, because of the absence of intrinsic roughness in atomically thin TMDs it has no role in charge scattering \cite{Bhoir2019}.
Coulomb scattering in 2D TMDs is caused by random charged impurities, located within the 2D layer or on its surfaces, and is the dominant scattering mechanism
at low temperatures as it is for graphene \cite{Hwang}.

Consider the scattering of CC on charged impurity centers
in a TMD monolayer placed on a substrate with a dielectric constant $\varepsilon_{s}$ and exposed to air on its upper side.
The scattering centers are distributed randomly in the plane of the monolayer, and across its thickness between the surfaces $z= \pm d/2$, where $d$ is the thickness of the monolayer. The assumption of a random nature of the distribution of impurity centers takes place when $d/\overline{r} \ll 1$, where $\overline{r}=(\pi N_{d})^{-1/2}$ is the mean distance between the impurity centers of concentration $N_{d}$.
Thus, for a monolayer of MoS$_{2}$ ($d=6.5$ \AA \cite{Radisavljevic}) the above inequality takes place at a concentration value of $N_{d} \ll N_{0}=d^{-2} \simeq 2.4 \cdot 10^{14}cm^{-2}$, which significantly exceeds the concentrations $N_{d} \simeq 5 \cdot 10^{11} cm^{-2}$ corresponding to strong doping \cite{Kaasbjerg}.
It should be also noted that the condition $k_{F}d \propto n^{1/2} d \ll 1$, where $k_{F}$ is the Fermi wave-number, is equivalent to the assumption that the scattering 
form factor of the impurity center is equal to unity \cite{Ando}.
For scatterings on Coulomb centers, the transport relaxation time of CC in the Born approximation, taking into account the valley degeneracy factor $g_{v}=2$ in TMDs \cite{Kaasbjerg}, is given by the expression \cite{Ando, Davies1997}:
\begin{equation}
    \frac{1}{\tau (E_{\vec{k}})}=\frac{4\pi N_{d}}{\hbar} \int \frac{d \vec{k}'}{(2\pi)^{2}}|V_{sc}(\vec{k},\vec{k}')|^{2}(1-\cos\vartheta)\delta (E_{\vec{k}}-E_{\vec{k}'}),
\end{equation}
where $\hbar$ is the reduced Planck's constant, $V_{sc}(\vec{k},\vec{k}')$ is the Fourier transform of the screened Coulomb potential of CC interaction with an impurity center, $\vec{k}$ is 2D wave-vector, $E_{\vec{k}}=\hbar^{2}\vec{k}^{2}/2m$ is kinetic energy and $m$ is the effective mass of the CC.
Let us denote the 2D Fourier transform of the energy of the Coulomb interaction of an electron with an impurity center in a vacuum by $V_{c}(\vec{q})$ \cite{Ando}, where $\vec{q}$ is a two-dimensional wave vector.
Taking into account both macroscopic nonlocal screening and screening by conduction electrons, the Fourier transform of the Coulomb interaction energy can be represented as:
\begin{equation}
    V_{sc}(\vec{q})=\frac{V_{c}(\vec{q})}{\varepsilon_{nl}(\vec{q})\varepsilon\vec(q)},
\end{equation}
where $\varepsilon_{nl}(\vec{q})$ defines the macroscopic nonlocal screening of a point charge in 2D dielectric layer and has the form:
\begin{equation}
   \varepsilon_{nl}(\vec{q}) = 1+r_{0}q,
\end{equation}
where the parameter $r_{0}=2\pi \alpha_{2D}$ is the characteristic length of 2D-electron screening dependent on the polarizability $\alpha_{2D}$ of the 2D dielectric layer \cite{Cudazzo1}. 
It should be noted, that when the 2D-layer is located between the media with dielectric constants $\varepsilon_{1}$ and $\varepsilon_{2}$, the parameter $r_{0}$ should be replaced by $r_{0}/\overline{\varepsilon}$, where $\overline{\varepsilon}=(\varepsilon_{1}+\varepsilon_{2})/2$ \cite{Durnev}.
The screening of the Coulomb center field by conduction electrons is described by the dielectric function of the 2D electron gas \cite{Ando}.
Taking into account dielectric surroundings of the monolayer and when $k_{F}d \ll 1$ (formally for $d \rightarrow 0$) \cite{Konar} the dielectric function can be presented in the form
\begin{equation}
    \varepsilon(\vec{q})=\overline{\varepsilon}\big[1+\frac{4}{qa_{B}} \Pi(q, T, \eta) \big],
\end{equation}
where $a_{B}=\hbar^{2}\overline{\varepsilon}/me^{2}$ is the effective Bohr radius and $\Pi(q, T, \eta)$ describes the $q - $dependency of the statistic polarization operator \cite{Ando}.

\begin{figure}
    \includegraphics[width=0.4\textwidth]{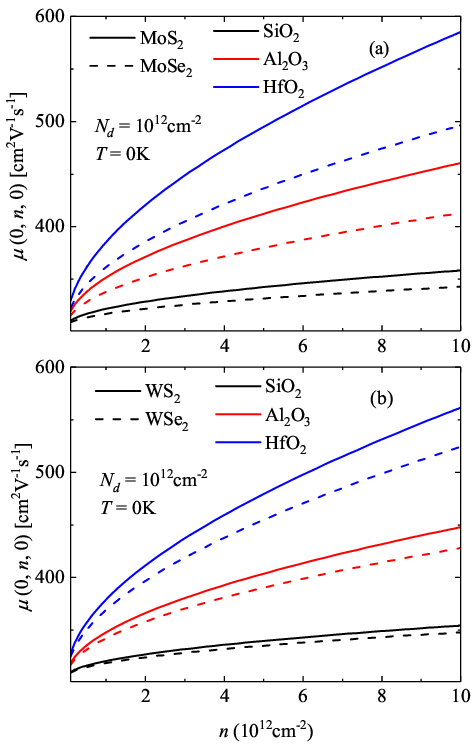}
    \vspace{-0.2cm}
    \caption{The dependencies of the mobility in TMD monolayers on electron concentration for $T=0$K and for various substrates in the case of $\delta = 0$: a. for MoS$_2$ and MoSe$_2$, b. for WS$_2$ and WSe$_2$.}
    \label{Enonint1}
\end{figure}
\begin{figure}
    \includegraphics[width=0.4\textwidth]{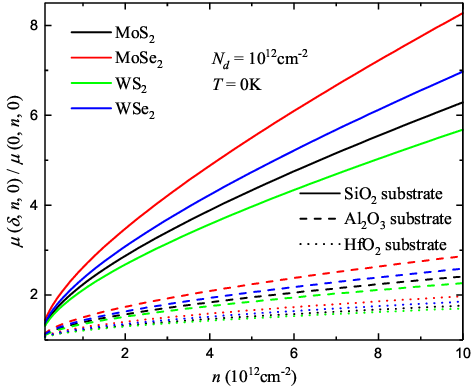}
    \vspace{-0.2cm}
    \caption{The dependencies of the mobility ratios in TMD monolayers on electron concentration for $T=0$K and for various substrates.}
    \label{Enonint1}
\end{figure}
\begin{figure*}
    \includegraphics[width=0.8\textwidth]{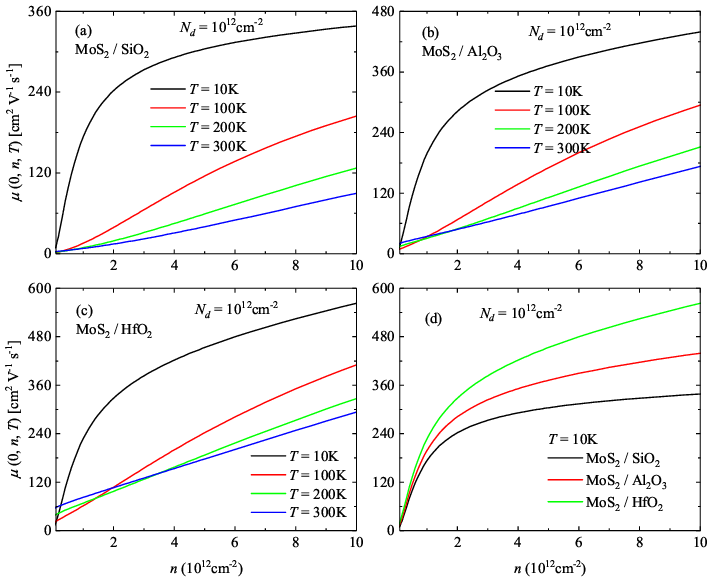}
    \vspace{-0.2cm}
    \caption{The dependencies of the mobility of TMD monolayers on electron concentration for different temperatures in case of $\delta=0$ ((a), (b), (c)). In (d) the electron mobility in MoS$_2$ TMD on different substrates is presented at $T=10$K.}
    \label{Enonint1}
\end{figure*}
\begin{figure*}
    \includegraphics[width=0.8\textwidth]{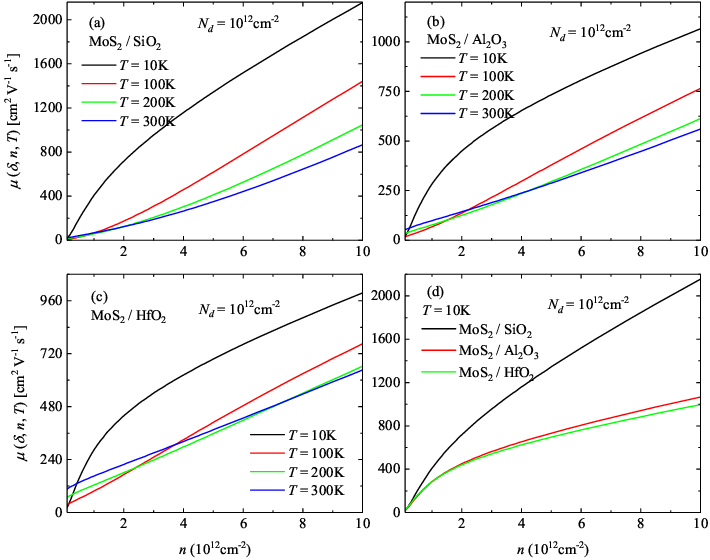}
    \vspace{-0.2cm}
    \caption{The dependencies of the mobility of TMDs monolayers on electron concentration for different temperatures in case of $\delta \neq 0$ ((a), (b), (c)). In (d) the electron mobility in MoS$_2$ TMD on different substrates is presented at $T=10$K.}
    \label{Enonint1}
\end{figure*}
\begin{figure}
    \includegraphics[width=0.4\textwidth]{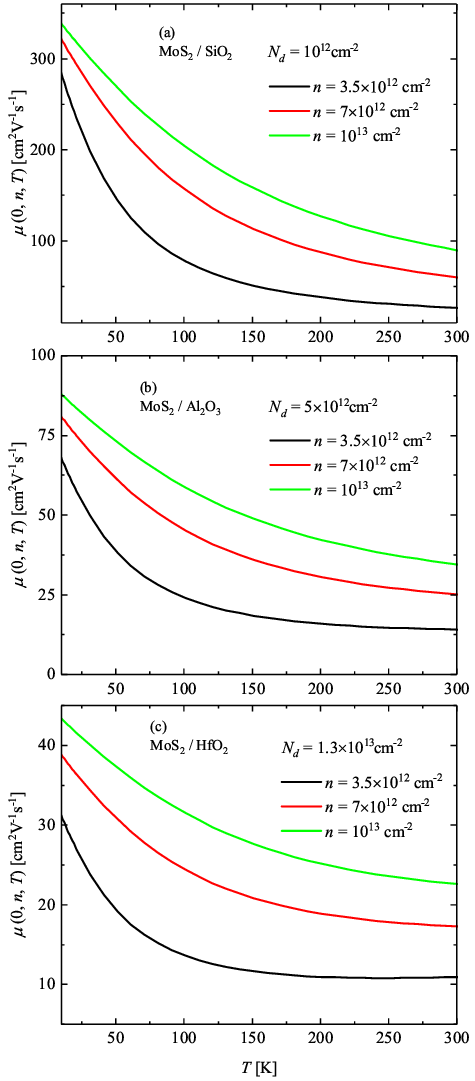}
    \vspace{-0.2cm}
    \caption{The dependencies of the mobility in MoS$_2$ monolayer on temperature for various substrates and concentrations of electrons and scattering centers in the case of $\delta = 0$.}
    \label{Enonint1}
\end{figure}
\begin{figure}
    \includegraphics[width=0.4\textwidth]{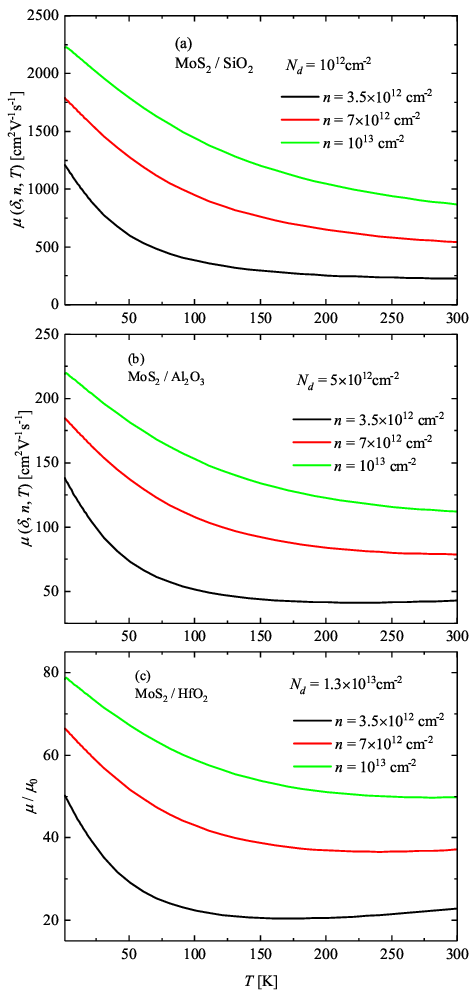}
    \vspace{-0.2cm}
    \caption{The dependencies of the mobility in MoS$_2$ monolayer on temperature for various substrates and concentrations of electrons and scattering centers in the case of $\delta \neq 0$.}
    \label{Enonint1}
\end{figure}
\begin{figure}
    \includegraphics[width=0.4\textwidth]{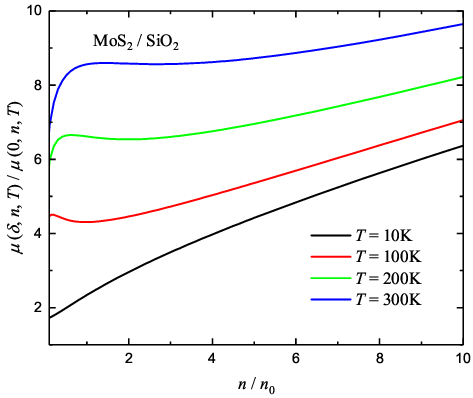}
    \vspace{-0.2cm}
    \caption{The dependencies of the mobility ratios calculated for MoS$_2$ on SiO$_2$ substrate with and without nonlocal scattering as functions of electron concentration at four different temperatures.}
    \label{Enonint1}
\end{figure}

For finite temperatures the function $\Pi(q, T, \eta)$ describing the $q$- and $T$ - dependencies of the static polarizability function is given by the formula \cite{Maldague1978, Ong2013}. 

\begin{equation}
    \Pi (q,T,\eta) = \int_{0}^{\infty} \frac{\Pi (q,0,E) dE}{4k_{B}T ch^{2}(\frac{\eta -E}{2k_{B}T})},
\end{equation}
where $\eta$ is the chemical potential of the system for given temperature, $E_{F}$ is the Fermi energy,
\begin{equation}
    \Pi (q,0, E_{F}) = 1- \sqrt{1-\left(\frac{2k_{F}}{q}\right)^{2}} \Theta (q-2k_{F}),
\end{equation}
where $\Theta (x)$ is the Heaviside step function, $k_{F}=(2\pi n/g_{v})^{1/2}=(\pi n)^{1/2}$, and $q=2k\sin(\theta/2)$ is the modulus of the scattering vector with $\theta$ being the scattering angle from the initial momentum $\hbar \vec{k}$ to the final momentum $\hbar \vec{k'}$.

After substituting expressions (2)-(6) into Eq. (1) the inverse transport relaxation time can be represented as:
\begin{equation}
    \frac{1}{\tau (E_{k})}=\frac{2 \pi \hbar N_{d}}{m}I(\delta,k),
\end{equation}
where 
\begin{eqnarray}
&& I(\delta,k) = \nonumber \\
&&  \int_{0}^{\pi}\frac{(1-cos \vartheta)d\vartheta}{\left(1+2\delta k a_{B}sin \frac{\vartheta}{2}\right)^{2}\left[k a_{B}sin \frac{\vartheta}{2}+2\Pi(q,T,\eta)\right]^{2}}.
\end{eqnarray}
In Eq. (8) $\delta$ is the characteristic parameter of the problem \cite{Mughnetsyan},
\begin{equation}
    \delta = \frac{r_{0}}{a_{B} \overline{\varepsilon}} = \frac{m}{m_{0}} \frac{r_{0}}{a_{B0} \overline{\varepsilon}^{2}},
\end{equation}
were $m_{0}$ is the free electron mass, and $a_{B0}=0.529$ \AA. The values of parameter $\delta$ for various TMDs and various substrates are given in Table I.
Note, that in \cite{Durnev} as a characteristic parameter of the problem $\delta^{-1}$ has been used.
It is obvious from Eqs. (7) and (8) that taking into account the new mechanism of the nonlocal screening of the scattering center (i.e. $r_{0} \neq 0$)
always results in the increase of transport relaxation time and hence of electron mobility.

The electron mobility is defined as follows:
\begin{equation}
    \mu = \frac{e}{m} \langle \tau (E_k) \rangle,
\end{equation}
where the transport relaxation mean time is given by the expression \cite{Ando}
\begin{equation}
\langle \tau (E_{k})  \rangle = \frac{\int d\vec{k} E_{k} \tau (E_{k})(-\partial f / \partial E_{k})}{\int d\vec{k} E_{k}(-\partial f / \partial E_{k})},
\end{equation}
$f \equiv f(E_{k})$ is the Fermi distribution function.
The chemical potential in Fermi distribution is defined from the normalization condition for 2D electron gas and is given as:
\begin{equation}
    \eta = k_{B}T \ln \left[ \exp \left( \frac{E_{F}}{k_{B}T} \right) - 1\right],
\end{equation}
where $E_{F} = \pi \hbar^{2} n /2m$ is the Fermi energy in the monolayer.

It is convinient to present Eq.(10) in a more compact form:
\begin{equation}
\langle \tau (E) \rangle = \frac{1}{E_{F}} \int_{0}^{\infty} \tau(E)E\Delta(E,\eta,T)dE,
\end{equation}
where 
\begin{equation}
   \Delta(E,\eta,T) = \frac{1}{4k_{B}T}ch^{-2}\left(\frac{E-\eta}{2k_{B}T}\right).
\end{equation}
In the limiting case of $T \to 0$K Eq. (14) goes over into the Dirac delta function $\delta (E-E_{F})$.
From Eqs. (10), (11), (7) we get for the mobility:
\begin{equation}
    \mu (\delta, n, T) = \mu_{0}\frac{n_{0}}{N_{d}E_{F}}\int_{0}^{\infty}E\Delta(E,\eta,T)I^{-1}(\delta,k)dE,
\end{equation}
where $\mu_{0} = e/2 \pi \hbar n_{0} = 241 cm^{2}/V \cdot s$,
and $n_{0} = 10^{12} cm^{-2}$.
\section{Results and Discussion}

1. Let us consider the electron mobility in the limiting case of $T=0$K.
From Eqs. (9)-(15) it follows that
\begin{equation}
    \mu (\delta, n, 0) = \mu_{0} \frac{N_{d}^{0}}{N_{d}} I^{-1}(\delta, k_{F}),
\end{equation}
and is inversely proportional to the concentration of scattering centers in the specimen $N_{d}$ (note that $N_{d}^{0}=n_0=10^{12}cm^{-2}$).
The electron mobility during scattering on Coulomb centers without taking into account the new scattering mechanism 
can be calculated using Eq. (16) for the value of $\delta = 0$.

In the figures below, electron concentration values vary in the range of $(10^{11}-10^{13})cm^{-2}$. At concentrations lower than $10^{11}cm^{-2}$, multiple scattering effects become significant, and the Boltzmann equation ceases to be valid. At concentrations greater than $10^{13}cm^{-2}$, higher-order electrical subbands begin to fill, and the transfer becomes multi-subband \cite{Ando}. Note also that below we bring the results of calculations only for MoS$_2$ because for the other TMDs the results are generic with once for MoS$_2$ (see Figs. 1 and 2).

Fig. 1 shows the dependence of the mobility in TMD monolayers on the electron concentration
for SiO$_2$ (black), Al$_2$O$_3$ (red), and HfO$_2$ (blue) substrates, respectively.
As can be seen from Fig. 1, all the curves increase monotonically with the increase of electron concentration, and the order of the curves does not change depending on the substrate.
The increase in mobility when moving from the SiO$_2$ substrate to ones with higher mean dielectric constants
is due to the known effect of dielectric screening of the media surrounding the monolayer \cite{Konar}.
Note also that the relatively close values of mobilities for different TMDs 
are due to the close values of the electron effective masses 
and the effective Bohr radii in them (see Table I).

To clarify the role of macroscopic nonlocal screening
in the electron mobility of a TMD monolayer, using Eq. (16), we will consider the ratio $\mu (\delta, n, 0)/\mu (0, n, 0)$.
In Fig. 2 the ratios of mobilities calculated with and without nonlocal scattering as function of electron concentration for TMD monolayers are presented
in the case of three various substrates and at $T=0$K.
As can be seen from the Fig. 2, the nonlocal screening always
leads to the increase of the mobility regardless of the dielectric environment of the monolayer.
In the case of, for example, a MoS$_2$ monolayer on a substrate of SiO$_2$,
this increase exceeds the mobility value, without taking into account the indicated scattering mechanism,
by more than 5.5 times for electron concentration of $n = 5 \cdot 10^{12} cm^{-2}$.
When moving to substrates with higher values 
of the dielectric constant, the effect of nonlinear screening
on the mobility in the monolayer weakens. Such a behavior is caused by the fact that in contrast to "high-k" effect taking place for Coulomb scattering of CC on charged impurity centers with increasing in the dielectric constant of environment its influence on the mobility of charge carriers is decreased. This behavior of the ratio of mobilities on dielectric constant of environment is caused by an essential dependence of the nonlocal screening parameter on the dielectric constant of environment (see also Table I).  
Note that with the increase of the electron concentration
the effect of nonlinear screening on the mobility increases.
In the case of a free standing TMD monolayer the calculations show that because of the relatively large values of the parameter $\delta$ the mobility values increase more than an order of magnitude (e.g. for concentration $n=5 \cdot 10^{12} cm^{-2}$ for MoSe$_{2}$ $\mu(n,\delta,0)/\mu(0,n,0) \simeq 17.5$).

In Fig. 3 the dependencies of electron mobility on the concentration are presented for different temperatures without taking into account the new scattering mechanism. It is evident that at any fixed temperature and for any substrate, the mobility is a monotonically increasing function of the electron concentration. Moreover, the temperature value for a fixed substrate affects the mobility value: the lower the temperature, the higher the mobility. Thus, for a $\text{MoS}_2$ monolayer on a $\text{HfO}_2$ substrate at $n=5\times10^{12}cm^{-2}$ the ratio of the mobility at $T=10K, 100K, 200K$ and $300K$ to the mobility at $T=0K$ (see Fig. 1, (a)) is 0.83, 0.56, 0.34 and 0.32, respectively. It also follows from Fig. 3, (a)-(c), that at $T=100K$ the mobility, compared to its value at $T=10K$ (at $n=5\times10^{12}cm^{-2}$), decreases approximately by half, and at $T=200K$ and $T=300K$ it is significantly less, while the dependence of mobility on concentration is practically linear. This mobility behavior is shaped by two factors that influence the behavior of the electron gas \cite{Ma2014}. The electron gas is less polarizable at high temperatures and low concentration. Polarizability is caused by the spatial redistribution of the electron gas induced by Coulomb potential, thus it is proportional to $n$. As temperature increases, the thermal energy randomizes the electron velocities, accelerating the transition of the electron system back into equilibrium distribution, consequently weakening the polarization. The decrease of polarizability reduces the free-carrier screening and, consequently, the mobility. For comparison in Fig.3 (d). the electron mobility in MoS$_2$ on different substrates is presented at $T=10$K.

Fig. 4 shows the dependencies of mobility on the electron concentration for different substrates ((a), (b), (c)) at four fixed temperatures taking into account the nonlocal scattering mechanism ($\delta\neq0$), and Fig. 4 (d) shows the mobility values at $T = 10K$ on three substrates. As follows from the figures, a significant increase in mobility occurs at $T = 10K$ in the case of the $\text{SiO}_2$ substrate. Thus, in comparison with the case of $\delta=0$ (Fig. 3 (a), black line), the ratio of the mobility values is 4.48 in the case of the $\text{SiO}_2$ substrate, 1.94 in the case of the $\text{Al}_2\text{O}_3$ substrate and 1.53 in the case of the $\text{HfO}_2$ substrate. With increasing temperature, this ratio increases significantly. Thus, at $T = 300K$, the ratio is 8.8 (on the $\text{SiO}_2$ substrate), 4.36 on the $\text{Al}_2\text{O}_3$ substrate and 2.1 on the $\text{HfO}_2$ substrate. Such a behavior of the curves is the result of competition between the polarizability of the electron gas and the tendency to reach an equilibrium state with increasing temperature.

The dependencies of mobility on temperature for different substrates ((a), (b), (c)) and at fixed concentration values are shown in Fig. 5 in the case of $\delta = 0$ At a fixed electron concentration, changing the substrate from $\text{SiO}_2$ (a) to $\text{Al}_{2}\text{O}_3$ (b) or $\text{HfO}_2$ (c), i.e., with higher values of dielectric constants, leads to an increase in mobility, which, apparently, can be explained by the “high-k” effect of dielectric screening \cite{Konar}.
Thus, at $T=100$ K, the ratio of mobilities on $\text{SiO}_2$, $\text{Al}_{2}\text{O}_3$, and $\text{HfO}_2$ substrates at a concentration of $n=10^{13} cm^{-2}$ is $1:1.43:2$, respectively. At a concentration of $n=3.5 \cdot 10^{13} cm^{-2}$ (a, black line), the mobility decreases monotonically with increasing temperature, and the rate of drop monotonically decreases. This behavior of mobility occurs for all substrates.
As the electron concentration increases, the region of relatively rapid mobility growth shifts toward lower temperatures. As can be seen from Fig. 5, (c), at $n=3.5 \cdot 10^{13} cm^{-2}$ (black line), the mobility changes slightly in the range of $100-300 K$, but changes significantly in the range of $T<100 K$. This behavior of mobility is explained by the fact that as the electron concentration increases, the screening effect begins to prevail, so a decrease in temperature significantly affects mobility behavior.

Fig. 6 shows the dependencies of mobility on temperature for three values of electron concentration and for three substrates ((a), (b), (c)) taking into account the new scattering mechanism.
From a comparison of the curves in Fig. 6 (a), (b), (c) with the ones in Fig. 5 (a), (b), (c) it follows that the nonlocal screening leads to a sharp increase in mobility for all concentrations and for all substrates.
Thus, at a concentration of $n = 10^{13} cm^{-2}$ for the $\text{SiO}_2$ substrate (Fig. 5 (a) and Fig. 6 (a), green lines) the mobility ratio is $6.2$, for the $\text{Al}_{2}\text{O}_3$ substrate is $2.56$, and for the $\text{HfO}_2$ substrate is  $1.77$.

The significantly greater increase in the case of the $\text{SiO}_2$ substrate is explained by the larger value of the nonlocal screening parameter: $r_0 = 41.45$ \AA and $\delta = 4.83$
(see Table I). As the concentration decreases, a very weakly expressed minimum appears (black line in Fig. 6 (b)) at $T=200 K$, and upon moving to the $\text{HfO}_2$ substrate (Fig. 6 (c), black line), the minimum becomes quite noticeable and converges at the point $T=165$K. Such a behavior of $\mu (\delta, n, T)$ can apparently be explained by competing tendencies of screening, which weakens with decreasing electron concentration, and the "high-k" effect caused by the $\text{HfO}_2$ substrate.

Fig. 7 presents the mobility ratios calculated for MoS$_2$ monolayer on SiO$_2$ substrate with and without nonlocal screening as functions of electron concentration at four different temperatures. It is evident that for all curves, starting with concentrations of $n \simeq 4n_0$, there is a nearly linear increase with concentration, while the rate of change in the mobility ratios decreases with increasing temperature. At concentrations of $n < 4n_0$, a greater increase in the mobility ratio corresponds to curves with higher temperatures, which is also explained by the weakening of effective screening (at a given concentration) with increasing temperature.

Let us compare some of our results for the mobility obtained for electron scattering on Coulomb centers in a $\text{MoS}_2$ monolayer with existing results, although the variety of model representations and approximations used, significantly complicates an unambiguous interpretation of the results of comparison.

Let us consider the experimental results for determining the mobility given in \cite{Zhang2023}, presented in graphical form. In the case of an $\text{Al}_2\text{O}_3$ substrate (\cite{Zhang2023}, Fig. 1, (g)), with an electron concentration $n=10^{13} \text{cm}^{-2}$ and Coulomb center concentration $N_d=5\times10^{12}\text{cm}^{-2}$, at $T = 100$K from the graph for $\mu_{Cl}$ we find the mobility value $\mu_{Cl} \simeq 100\text{cm}^2\text{V}^{-1}\text{s}^{-1}$, and at $T = 300$K $\mu_{Cl} \simeq 75\text{cm}^2\text{V}^{-1}\text{s}^{-1}$. Our calculation using Eq. (15) at $\delta=0$ (Fig. 5, (b)) gives: $\mu\simeq25\text{cm}^2\text{V}^{-1}\text{s}^{-1}$ at T=100K and $\mu\simeq12.8\text{cm}^2\text{V}^{-1}\text{s}^{-1}$ at T=300K. At $\delta \neq0$ we get: $\mu\simeq50\text{cm}^2\text{V}^{-1}\text{s}^{-1}$ at T=100K and $\mu\simeq38\text{cm}^2\text{V}^{-1}\text{s}^{-1}$ at T=300K (Fig. 6, (b)).
In the case of the $\text{HfO}_2$ substrate (\cite{Zhang2023}, Fig.1, (h)), with values $n=7\times10^{12}\text{cm}^{-2}$ and $N_d=1.3\times10^{13}\text{cm}^{-2}$ for $\mu_{Cl}$ we have: $\mu_{Cl}\simeq60\text{cm}^2\text{V}^{-1}\text{s}^{-1}$ at $T=100$K and $\mu_{Cl}\simeq35\text{cm}^2\text{V}^{-1}\text{s}^{-1}$ at $T=300$K. Calculation using Eq. (15) with $\delta=0$ (Fig. 5, (c)) yields: $\mu\simeq24\text{cm}^2\text{V}^{-1}\text{s}^{-1}$ at $T=100$K and $\mu\simeq17.5\text{cm}^2\text{V}^{-1}\text{s}^{-1}$ at $T=300$K. With $\delta\neq0$, we get: $\mu\simeq43\text{cm}^2\text{V}^{-1}\text{s}^{-1}$ at $T=100$K and $\mu\simeq36\text{cm}^2\text{V}^{-1}\text{s}^{-1}$ at $T=300$K (Fig. 6, (c)).

From the comparison of the presented data, it follows that the mobility values calculated taking into account nonlocal screening ($\delta\neq0$) are closer to the mobility values given in [27] than the mobility values obtained without taking nonlocal screening into account.

\section{Conclusion}

In this paper we have provided a theoretical framework for electron transport in TMD monolayers that incorporates the two-dimensional nonlocal screening of Coulomb impurities, originally proposed by Cudazzo \textit{et al.}~\cite{Cudazzo1}. In contrast to the conventional dielectric approximation, this treatment accounts for the wave-vector dependence of the dielectric function, thereby capturing the intrinsic screening properties of atomically thin layers. Our results reveal that in the electron concentration range 
$n \sim 10^{11}\mathrm{cm^{-2}} -  10^{13}\,\mathrm{cm^{-2}}$, 
the inclusion of nonlocal screening enhances the mobility caused by Coulomb scattering by several times. The magnitude of this effect depends strongly on the dielectric environment: the substrates such as SiO$_2$ or free-standing monolayer exhibit the largest enhancements, whereas "high-$k$" dielectric substrates (e.g., HfO$_2$, Al$_2$O$_3$) substantially reduce the improvement due to the strong dependence of the screening parameter $\delta$ on the mean dielectric constant.
We also find that the role of nonlocal screening persists at finite temperatures. Although the mobility decreases with increasing temperature due to weakening of free-carrier screening, the enhancement introduced by the nonlocal mechanism actually grows $6$--$9$ times at room temperature on SiO$_2$. To summarize, the obtained results highlight the necessity of including nonlocal dielectric screening in theoretical models of mobility in TMDs, both for accurate comparison with experiments and for guiding the design of high-performance devices such as FETs. To compare the mobility values of TMDs with experimental data, one should calculate the temperature dependence of mobility considering the new scattering mechanism and compare it both with scattering by structural defects and with the dominant phonon scattering mechanism at intermediate and high temperatures.

\section{Acknowledgements}
\label{acknowledgements}
This work was financially supported by the Armenian State Committee of Science (grants No 24LCG-1C004, No 24WS-1C040 and No 21AG‐1C048).

\end{document}